\documentclass[journal]{IEEEtran}
\usepackage{color}
\usepackage{cite}

\ifCLASSINFOpdf
\else
  \usepackage[dvips]{graphicx}
\fi
\usepackage[cmex10]{amsmath}

\usepackage{cases}

\begin{document}
%
\title{Readout system with on-board demodulation for CMB polarization experiments 
using coherent polarimeter arrays}
%
%
%

\author{Koji~Ishidoshiro,~Makoto~Nagai,~Takeo~Higuchi,~Masaya~Hasegawa,
~Masashi~Hazumi,~Masahiro~Ikeno,~Osamu~Tajima,~Manobu~Tanaka,~Tomohisa~Uchida
\thanks{Manuscript received 29 November 2011: Revised 10 February 2012.}
\thanks{Authors are with High Energy Accelerator Research Organization~(KEK) Tsukuba, Japan 305-0801 (E-mail koji@post.kek.jp)}%

}

\maketitle

\begin{abstract}
$B$-modes are special patterns in cosmic microwave background (CMB) polarization. 
Degree-scale $B$-modes are smoking-gun signatures of primordial gravitational waves. 
The generic strategy of CMB polarization experiments is to employ 
a large number of polarimeters for improving the statistics. 
The Q/U Imaging ExperimenT-II (QUIET-II) has been proposed to detect 
$B$-modes by using the world's largest coherent polarimeter array (2,000 channels). 
A unique detection technique using QUIET's polarimeters, 
which comprises a modulation/demodulation scheme, enables us to directly extract polarization signals. 
The extracted signals are free from unpolarized components and intrinsic $1/f$ noise. 
We developed a data readout system for the QUIET-II experiment. 
We employed a ``master'' clock strategy, on-board demodulation functions, and end-to-end Ethernet connections for logical simplicity and 
high-density compact electronics for physical compactness. 
A clock module acts as a single master 
and guarantees phase matching between the modulation by the polarimeters and the on-board demodulation by ADC modules.  
Each ADC module has 64 ADC chips in the VME-6U single slot size. 
Both modules have hardware processors for Ethernet TCP/UDP. 
All these modules and control computers are connected via end-to-end Ethernet. 
Physical compactness and logical simplicity enable us to easily handle a large number of polarimeters, 
while maintaining quality of the $B$-mode experiments. 
The developed electronics (the clock modules and the ADC modules) fulfill these requirements. 
Tests with a setup similar to that of the real experiment proved that the system works appropriately. 
The performance of all system components is validated to be suitable for $B$-mode measurements. 
\end{abstract}


\IEEEpeerreviewmaketitle

\section{Introduction}
\PARstart{D}{etection} of primordial gravitational waves could provide 
a new and unique window on the very early universe~\cite{dodelson}. 
Although various approaches for detecting them \cite{ligo,toba,cassini}, 
the most promising approach is the measurement of $B$-modes, 
which are odd-parity patterns in cosmic microwave background (CMB) polarization. 
Because $B$-modes are very faint ($<100$~nK), 
it is important to detect a large number of CMB photons for achieving sufficient sensitivity. 
The Q/U Imaging ExperimenT (QUIET) is a ground-based experiment that 
aims to detect the $B$-modes in the Atacama Desert in Chile, 
which is $5,080$~m above sea level.
At the initial phase of the QUIET experiment (QUIET-I), 
we observed CMB polarization in the 95 (43)~GHz bands with 360 (72) channels of coherent polarimeter elements (polarimeter array). 
The obtained upper bounds for the $B$-modes are one of the most stringent limits to date; 
however they are still limited by statistical errors \cite{QUIET}. 
An upgrade of the experiment (hereafter referred to as QUIET-II) has been 
proposed to achieve better sensitivity with a larger polarimeter array. 
Its primary goal is to detect $B$-modes at $r \approx 0.01$ by using 2,000 channels 
composed of 500 elements of four-output polarimeters, 
where $r$ is the tensor-to-scalar ratio 
that indicates the intensity of the primordial gravitational waves \cite{r}.

We developed a readout system for the QUIET-II experiments. 
For handling the QUIET-II polarimeter array, we employed a master clock strategy, on-board demodulation functions, and end-to-end Ethernet connections for logical simplicity 
and developed high-density electronics to achieve physical compactness. 
The master clock strategy guarantees phase matching between modulation and demodulation.
Since polarization signals are modulated to suppress polarimeter noise and unpolarized components, 
phase matching and demodulation functions are essential requirements for the system. 
The on-board demodulation has the advantage of reducing the load of data transfer. 
Furthermore, physical compactness is also important to attach the system on a telescope mount to avoid picking up unexpected noise 
on the cables between the polarimeter array and the readout system. 
In this paper, we describe the system design and prototype tests.


\section{Extraction of the polarization signal}

\subsection{QUIET polarimeter}
A single QUIET polarimeter enables us to directly measure the Stokes $Q$ and $U$ parameters 
that characterize liner polarization. 
The polarimeters are polarization-sensitive coherent detectors using high electron-mobility transistor (HEMT) ampliﬁers, 
which have been widely used in CMB experiments \cite{CAPMAP,WMAP,PLANK}. 
They are suitable for detecting CMB polarization below 100~GHz. 
The polarimeter consists of a septum polarizer and strip-line-coupled monolithic microwave integrated circuit devices (Fig.~\ref{fig-module}). 
Input radiation is split into right and left circular polarizations ($E_R$ and $E_L$, respectively) by the spectrum polarizer. 
In the devices, the two polarization signals are modulated and coupled and then the power of the coupled signals is converted to analog outputs. 
In the polarimeter outputs, the Stokes parameters appear as the modulation components. 

The QUIET collaboration developed a polarimeter-on-a-chip, which replaces waveguide-block components with the devices \cite{QUIET-module}. 
The resulting package has a footprint size of 2.5~cm $\times$ 2.5~cm (95~GHz band) and 5~cm $\times$ 5~cm (43~GHz band) in QUIET-I. 
This breakthrough technology enabled us to build a large polarimeter array. 

\subsection{Modulation/demodulation scheme}\label{sec-demod}
\begin{figure}
\centering
 \includegraphics[width=8cm]{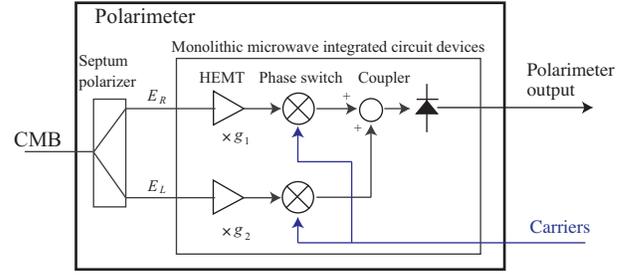}
 \caption{Block diagram of the QUIET polarimeter. 
It consists of a septum polarizer and strip-line-coupled monolithic microwave integrated circuit devices. 
The input radiation is split into right and left circular polarizations ($E_R$ and $E_L$, respectively) by the septum polarizer. 
They are amplified by the HEMTs. After applying the modulation on the phase switches, they are recombined by couplers. 
The couplers have four output ports. The power of each is detected by a Schottky diode. 
The four diode outputs have the modulation components related to $+Q$, $-Q$, $+U$, or $-U$ Stokes parameters, respectively. 
Three of the four diodes are not shown in this diagram. 
}\label{fig-module}
\end{figure}

An HEMT amplifier is known to have a $1/f$ noise whose knee frequency is high approximately 1~kHz. 
A modulation/demodulation scheme is employed to extract the polarization signal without any contamination 
from the $1/f$ noise and unpolarized components~\cite{QUIET-module}. 
The block diagram of the modulation is shown in Fig.~\ref{fig-module}. 
The input $E_R$ and $E_L$ to the devices are amplified by HEMTs. 
Then, the amplified polarizations are modulated by phase switches, 
each of which has two microwave paths. 
Path selection is controlled by p-intrinsic-n diodes. 
The difference among these path lengths corresponds to a half-wavelength of the measured CMB. 
The phase switch varies the phase of the CMB signal by $180^\circ$, 
which simply flips the sign of the CMB signal, and there is no sign flip for noises. 
Applying carrier clocks to the diodes offers periodic phase (sign) modulation. 
The two amplified and modulated polarizations are recombined by couplers. 
Four voltage-biased square-law Schottky diodes detect the power of the coupled signals. 

Suppose we implement the modulation only for one side, e.g., $E_R$, then 
the polarimeter outputs (or diode receiving power) can be expressed as follows (see Appendix):
\begin{equation}\label{eq-Dt0}
 D(t) = c_1(t)P + c_1(t)^2R + |c_1(t)|^2 N_1(t), 
\end{equation}
where $P$ is proportional to $g_1g_2 Q$, $-g_1g_2Q$, $g_1g_2U$, and $-g_1g_2U$ for each diode, 
$R$ corresponds to the power of the right circular polarization, i.e., ($g_1^2 E_R^2$).
$N_1(t)$ is the intrinsic noise of the HEMT amplifier on the side of $E_R$, 
$g_1$ and $g_2$ are gains of the HEMT amplifier, and 
$c_1(t)$ is the primary modulation term. 
We can set $c_1(t_{2p})=+1$ and $c_1(t_{2p+1})=-1$ in case of ideal modulation. 
Here, $p$ is the normalized time index with a half period of $c_1(t)$ and 
$D(t_p)$ is the averaged value during the $p^{\rm th}$ half period.  
Polarization $P$ can be extracted from the difference of each state, i.e., $\Delta D= D(t_{2i})-D(t_{2i+1})$. 
In Eq.~(\ref{eq-Dt0}), we drop the terms that are related to the noise from another HEMT amplifier.  

The modulation by the real polarimeter is not perfect, e.g., 
$c_1(t_{2p})=+1$ and $c_1(t_{2p+1})=-(1-\epsilon_1)$, where $\epsilon_1$ is a small nonzero value. 
As a result, residual terms that are related to the $R$ and $N_1(t)$ appear. 
These residuals can be eliminated by the secondary modulation for another side \cite{yuji}.
With the secondary modulation for $E_L$, the polarimeter output can be described as follows:
\setlength{\arraycolsep}{0.0em}
\begin{eqnarray}\label{eq-Dt}
D(t)&{}={}& c_1(t)c_2(t)P + c_1(t)^2R + c_2(t)^2L \IEEEnonumber \\
&& + |c_1(t)|^2 N_1(t) + |c_2(t)|^2 N_2(t), 
\end{eqnarray}
\setlength{\arraycolsep}{5pt}
where $L$ and $N_2$ indicate the power of the left circular polarization ($\propto g_2^2E_L^2$) and the intrinsic noise on the side of $E_L$, respectively. 
Here, $c_2(t)$ is similar to $c_1(t)$, i.e., 
$c_2(t_{2q})=+1$ and $c_1(t_{2q+1})=-(1-\epsilon_2)$, 
where $\epsilon_2$ is a small nonzero value that indicates imperfection of the second modulation 
and $q$ is the normalized time index with the half period of the second modulation. 
This time period is longer than that of the primary modulation. 
Only the first term in Eq.~(\ref{eq-Dt}) has the phase information of both modulations. 
Therefore, the series of difference allows us to extract the polarization ($P$) without any contamination. 
This scheme is called demodulation. 
We implement this function in the readout system.  

\section{System design}\label{sec-design}
\subsection{System structure}
\begin{figure}[!t]
\centering
\includegraphics[width=8cm]{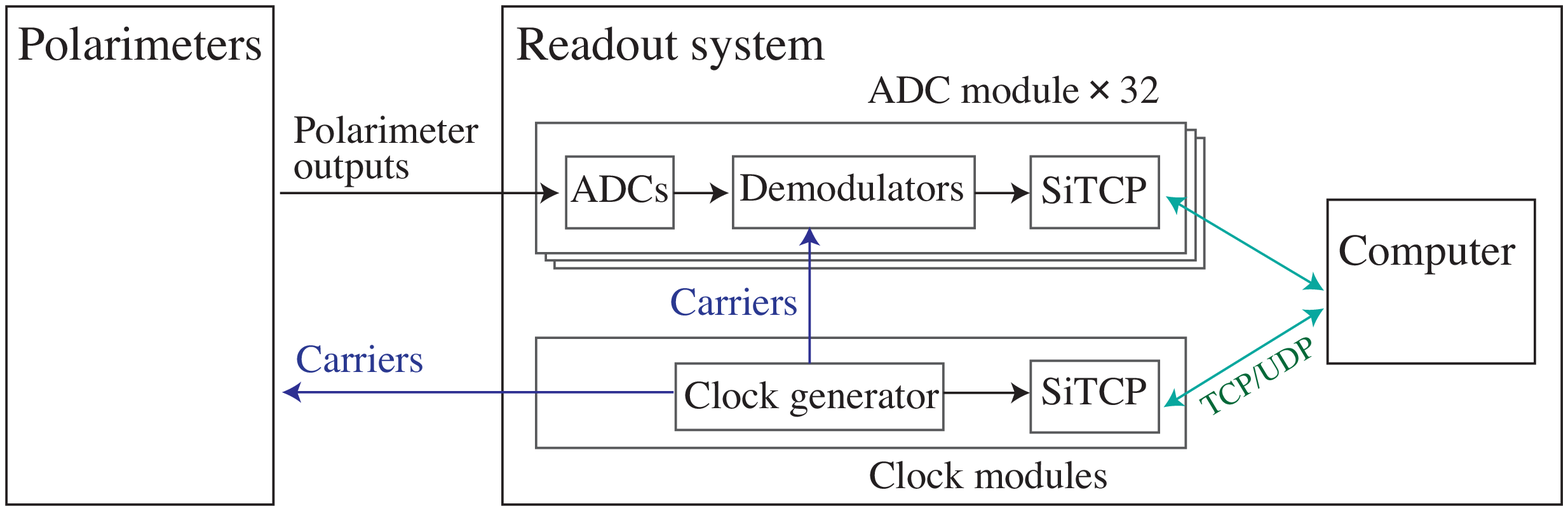}
\caption{Conceptual block diagram of the readout system. }\label{fig-overview}
\end{figure}

\begin{figure}
\centering
 \includegraphics[width=8cm]{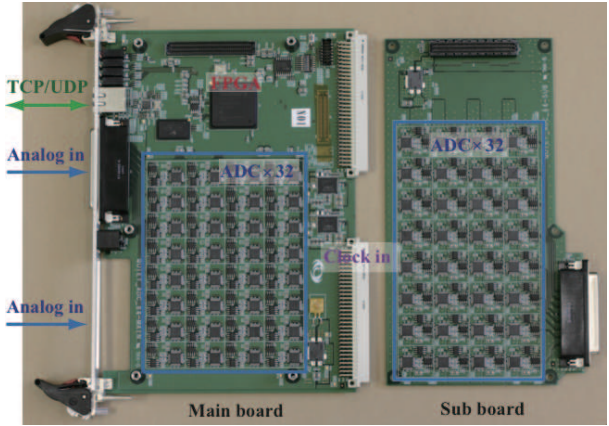}
 \caption{Photograph of an ADC module that consists of a main board and a sub-board. 
 Each board houses 32 ADC chips (AD7674) and 32 buffer amplifiers (AD8028). 
 A single FPGA (XC6SLX150) is mounted on the main board for the control of ADC chips, 
 demodulation functions, and SiTCP. 
}\label{fig-photoadc}
\end{figure}

The system structure is shown in Fig.~\ref{fig-overview}. 
The logical simplicity of our structure is possible because of the master clock strategy, on-board demodulation functions, and end-to-end Ethernet connections.  
Clocks acts as carriers to indicate phase states for the modulation. 
They should be in-phase between the polarimeter and the readout system. 
To retain phase matching, we employ clock modules that work as a single ``master.'' 
The clock modules generate well-matched carrier clocks and distribute them to the polarimeter array and analog-to-digital converter (ADC) modules. 
Each ADC module performs digitization and on-board demodulation for 64 channels. 
Having a sub-board in addition to a main board, 
a single ADC module houses 64 ADC chips (32 chips each on the main board and the sub-board) in a versa module eurocard~(VME) 6U single slot size (Fig.~\ref{fig-photoadc}).
All circuits for the ADC chip control and on-board demodulation are implemented in a single field programmable gate array (FPGA). 
Such high-density and compact modules are suitable for attaching to the telescope mount, 
which is a basic scheme to suppress the noise picked up on the cable between the polarimeters and the ADC modules. 

The demodulated data in the ADC modules are transmitted to a readout computer via Ethernet transmission control protocol (TCP)~\cite{nagai}. 
To control both modules from the computer, user datagram protocol (UDP) is also used.  
TCP and UDP in both modules are implemented by a hardware-based processor called SiTCP \cite{SiTCP}. 
Although to handle TCP and UDP, an additional FPGA or CPU is usually required, 
we do not need them because of the advantages of SiTCP. 

\subsection{Specifications of the ADC chip}\label{sec-spec}
Analog-to-digital (A/D) conversion is synchronized with an A/D clock. 
Its sampling rate ($f_{\rm A/D}$) is determined to avoid ringing spikes 
during the phase flip. 
The ringing spikes appear for 15--20~$\mu$sec. 
We can avoid the effects of the spikes by masking these periods. 
To optimize the masking region with a precision better than 10\%, $f_{\rm A/D}>666$~kHz is required.  

The system should be able to measure signals ranging from the CMB temperature (2.7~K) 
to room temperature (300~K). 
Suppose the responsivity of the polarimeter is $\approx 10$~mV/K, 
which is the highest among QUIET-I polarimeters, 
then the ADC dynamic range $V_{\rm d}$ should be at least $V_{\rm d}>3,000$~mV. 

The noise level of the QUIET-II polarimeter is expected to be $\approx 10^{-3}$~mV/Hz$^{1/2}$,  
assuming a sensitivity of 500~$\mu$K$\sqrt{\rm s}$/channel (250~$\mu$K$\sqrt{\rm s}$/polarimeter).
Considering the individual differences among the polarimeters, 
the lowest noise level may be $\approx 3 \times 10^{-4}$~mV/Hz$^{1/2}$. 
With a safety factor of $10$, 
we have to determine the ADC resolution $R$ to fulfill the following equation: 
\begin{equation}\label{eq-adc}
 \frac{\Delta_{\rm A/D}}{\sqrt{6 f_{\rm A/D}}}~~{\rm mV/Hz}^{1/2} <  3 \times 10^{-5}~~{\rm mV/Hz}^{1/2},  
\end{equation} 
where $\Delta_{\rm A/D}$ is the least significant bit defined as $\Delta_{\rm A/D} \equiv V_{\rm d}/2^R$. 
The left-hand side of Eq.~({\ref{eq-adc})} is the ADC quantization noise estimated 
from the root mean square of the quantization noise ($\Delta_{\rm A/D}/\sqrt{12}$) divided by the square of the band width $f_{\rm A/D}$
and the amplification of it by $\sqrt{2}$ to obtain a one-sided spectrum. 

\begin{figure}[!t]
\centering
 \includegraphics[width=9cm]{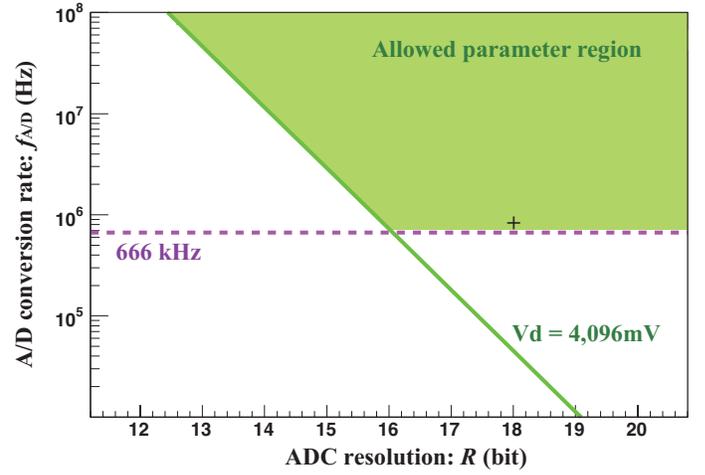}
 \caption{Allowed parameter region for the ADC sampling speed and resolution space under the condition $V_{\rm d}=4,096$~mV. 
 The solid lines are from Eq.~(\ref{eq-adc}) and the dashed lines denote the ringing spike avoidance ($f_{\rm A/D}=666$~kHz). 
 The allowed region is colored in green. 
 The cross mark indicates the ADC specification used in the readout system. 
}\label{fig-adc-req}
\end{figure}

For the ADC chip parameters, the allowed region based on the above requirements 
under the condition $V_{\rm d}=4,096$~mV is shown in Fig.~\ref{fig-adc-req}. 
We use a commercial ADC chip, AD7674~\cite{ADC}, that has an 18~bit resolution with $f_{\rm A/D}=800$~kHz and $V_{\rm d}=4,096$~mV.  
For the ADC chip control as well as the on-board demodulation, we use a system clock with $f_{\rm sys} = 40$~MHz. 

This frequency is selected such that it is sufficiently high to operate the ADC chips in the serial interface mode but 
sufficiently low to allow easy handling. 
In principle, a slower system clock is possible when we select the parallel interface mode. 
However, such an interface requires more electronic circuitry and a larger FPGA on board, 
which is contrary to the objectives of designing a high-density compact module. 

\subsection{Modulation/demodulation frequencies and data recording rate}
The demodulated data can be downsampled at the data recording rate ($f_{\rm record}$) without any loss of CMB information. 
This rate determines the sampling of the sky. 
The mount for the QUIET-II experiment is desired whose scan speed $\dot{\theta}_{\rm scan}$ is $6^\circ$/s at the maximum 
with a full width at half maximum (FWHM) of $0.1^\circ$ for the angular resolution ($\Delta \theta$). 
The relationship $2\times \dot{\theta}_{\rm scan} \times 1/f_{\rm record}<\Delta \theta$ should be satisfied, 
where a factor of two is introduced by the sampling theorem. 
This relationship requires $f_{\rm record}>120$~Hz. 
Under this condition, we should also consider 
avoidance of the aliasing effect of AC power frequency and its harmonics.  

Two types of carrier clocks are sent to both polarimeters and ADC modules. 
The frequency of the primary carrier $f_{\rm c1}$ must be higher than the knee frequency of the $1/f$ noise,  
which mainly occurs from the HEMT amplifier of the polarimeter. 
Its typical knee frequency is $\approx 1$~kHz. 
To minimize the occurrence of ringing spikes, we minimize $f_{\rm c1}$. 
Therefore, $f_{\rm c1}=4$~kHz is the natural selection. 

The frequency of the secondary carrier $f_{\rm c2}$ should be downscaled 
to the frequency of $f_{\rm c1}$ by an even number, 
\begin{equation}
 f_{\rm c2}=\frac{f_{\rm c1}}{2n} ~~(n~{\rm is~an~integer~number}).
\end{equation}
This frequency is possibly lower than the knee frequency of the $1/f$ noise. 
The lower bound of $f_{\rm c2}$ is $f_{\rm record}$ and 
we choose $f_{\rm c2}=f_{\rm record}=125$~Hz. 

We summarize the specifications of the clocks in Table.~\ref{tab-clock}. 

\begin{table}
 \caption{Frequencies of clocks used in the system.}
 \label{tab-clock}
 \centering
 \begin{tabular}{lr}
 \hline 
 Clock & Frequency  \\ \hline 
 System clock ($f_{\rm sys}$) & 40~MHz \\  
 A/D clock ($f_{\rm A/D}$) & 800~kHz \\  
 Primary carrier ($f_{\rm c1}$) & 4~kHz  \\  
 Secondary carrier ($f_{\rm c2}$) & 125~Hz  \\  
 Recording rate ($f_{\rm record}$) & 125~Hz  \\  \hline 
 \end{tabular}
\end{table}

\section{Clock module}\label{sec-clock}
\begin{figure}
\centering
 \includegraphics[width=8cm]{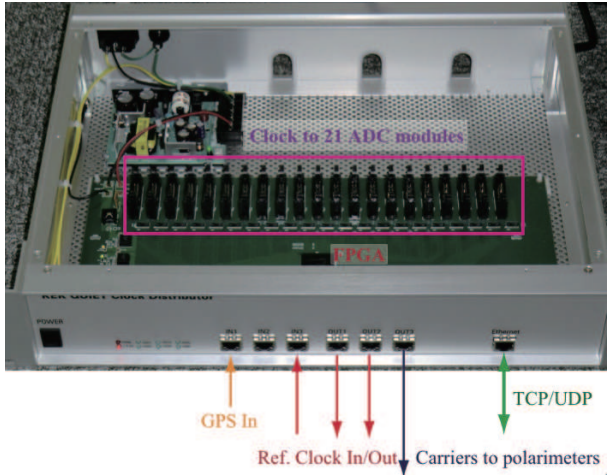}
 \caption{Photograph of the clock modules. 
 All clocks are generated and distributed from this module. }\label{fig-masterclock}
\end{figure}

We developed 19 inch 2U size clock modules (Fig.~\ref{fig-masterclock}).
A single module has seven RJ45 slots on the front panel and 21 flat cable (20 pins) connector slots on the surface of the board.
The rightmost RJ45 slot is used for communication with the computer via TCP/UDP. 
All other input/output signals are at the level of LVDS.
Furthermore, all functions are controlled by FPGA (XC6SLX75) in the module.
Its firmware is stored in a platform flash PROM (XCF32P).
The IP address is stored in an electrically erasable programmable read-only memory (AT93C46).

\begin{figure}[!t]
\centering
 \includegraphics[width=8cm]{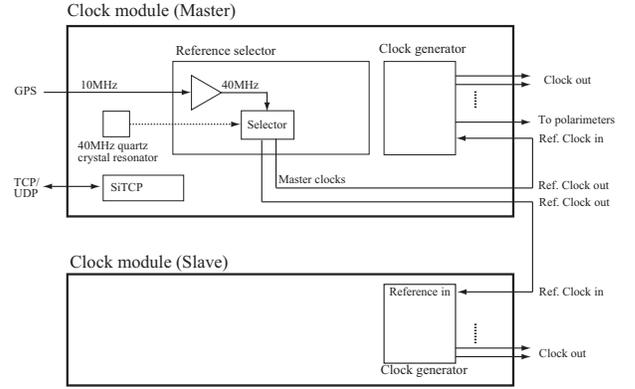}
 \caption{Functional diagram of the clock module. 
 The master and slave are identical modules. 
 The reference selector is only used in the master module. 
 All functions are implemented in a single FPGA. }\label{fig-clockchart}
\end{figure}

The module logic consists of two parts: 
a ``reference selector'' and a ``clock generator''   as shown in Fig.~\ref{fig-clockchart}. 
The selector performs the function of selecting a reference clock. 
The clock generator has two functions: generating the clocks on the basis of the reference clock 
and transmitting them to the ADC modules as well as the polarimeters.
Because of space limitations for the clock distribution slots, 
we use two clock modules to handle 2,000 channels: master and slave.
The selector generates a single reference, i.e., a ``master clock,'' for the master and slave.
Based on the reference ``master clock'', both modules generate the clocks and deliver them to each component.
Such a ``master clock'' strategy guarantees phase matching among the clocks. 

A GPS clock (10~MHz) is received via the leftmost RJ45 slot and 
is then multiplied into a 40~MHz clock in the module.
The internal quartz crystal resonator (KC7050C40.0000C3WE0) in the module also generates a 40~MHz clock. 
The reference selector chooses the master clock from these two options.
The internal clock is useful for laboratory testing when a GPS clock is not used. 
The master clock is duplicated and these two clocks are output via two RJ45 slots.
We found that the phase delay between them is 0.8~nsec, 
which is a suitable specification for the 40~MHz (25~nsec period) system clock. 

Each module has an RJ45 slot to receive the reference clock.
The use of the same cable length as that of the reference output slots in the master clock guarantees the 
same phase delay for the two reference clocks. 
In each module, the clock generator produces the clocks as listed in Table.~\ref{tab-clock}. 
The carrier clocks are distributed from the RJ45 slot to another board to control the carrier bias level of the polarimeters.
The last RJ45 input slot is reserved to test the timing of TCP/UDP.
A single flat cable is used for delivering the clocks to the single ADC module.
Two clock modules (the master and slave) can deliver the clocks to the 42 ADC modules 
(which have a combined capability of handling 2,688 channels).
The phase differences among the delivered carrier clocks were confirmed to be at most 2.4~nsec.
This is also a sufficient performance compared with the ADC sampling interval of 1.25~$\mu$sec~(=1/$f_{\rm A/D}$).

\section{ADC Module}\label{sec-adc}
The ADC module has a VME-6U single slot size (Fig.~\ref{fig-photoadc}). 
A single module consists of a main board and a sub-board. 
Each board has 32 ADC chips (AD7674) with 32 buffer amplifiers (AD8028) and a Dsub78 connector 
for 32 pairs of differential inputs with eight analog grounds. 
By using the sub-board, we successfully doubled the number of ADC chips 
compared with that in the previous generation ADC modules \cite{QUIET-ADC1,QUIET-ADC2}. 
A single module has the capability of handling 64 channels in total by using a single FPGA (XC6SLX150) on the main board.
We plan to use 32 modules to handle 2,000 inputs from the polarimeters. 

Each module works at +5~V and 3~A, i.e., 15~W. 
A VME crate supplies the power to the ADC modules. 
All clocks are provided from the clock module via VME P2 user-defined pins with a digital ground. 
The digital and analog grounds are connected via a zero-ohm connection at one location. 
%
An electrically erasable programmable read-only memory (AT93C46) records the IP address for the TCP/UDP communication.
The FPGA firmware is stored in a serial flash memory (M25P64-VMF).

\subsection{Firmware logic}\label{subsec-demod}

\begin{figure}[!t]
\centering
 \includegraphics[width=9cm]{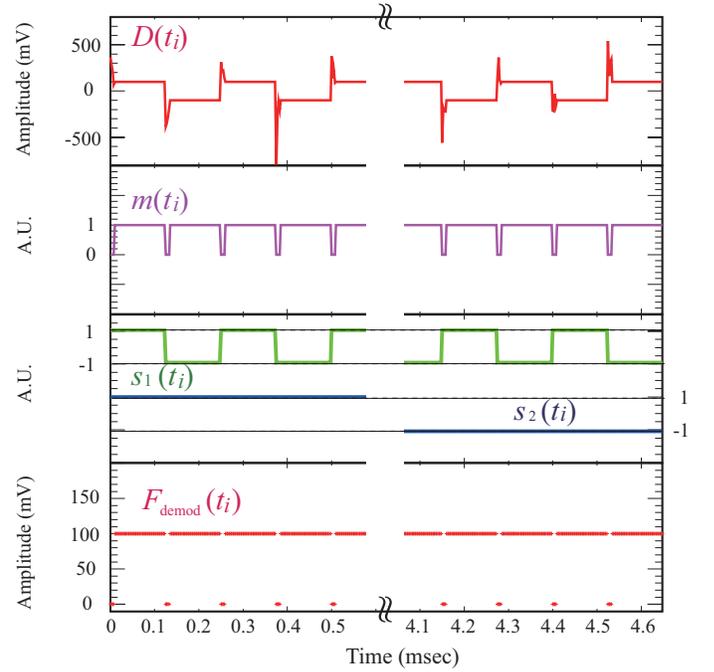}
 \caption{Schematics for the raw input at $f_{\rm A/D}=800$~kHz ($D(t_i)$), the mask for the ringing spikes ($m(t_i)$), the demodulation carriers ($s_1(t_i)$ and $s_2(t_i)$). 
 The demodulated stream at $f_{\rm A/D}=800$~kHz is obtained by multiplying the carriers to the masked stream. 
 The ringing spikes and the offset generated by the modulation are eliminated in the $F_{\rm demod}$ stream. 
}\label{fig-demod-chart}
\end{figure}

The polarimeter output includes ringing spikes induced by the phase flip, which must be masked. 
The top of Fig.~\ref{fig-demod-chart} shows a schematic example of the digitized polarimeter output stream $D(t_i)$ with the ringing spikes,  
where $i$ is the time index of the digitization rate ($f_{\rm A/D}=800$~kHz). 
Digitized ringing spikes are masked by multiplying a masking function $m(t_i)$ which has the value $0$ when ``mask''=true and $1$ when ``mask''=false 
(see an example in Fig.~\ref{fig-demod-chart}).  
For studying the ringing spikes, 64 raw digitized streams can be recorded during 300~$\mu$sec, 
which is limited by the size of the internal memory of the FPGA. 
We nominally mask one sample before and 13 samples after the phase change. 
The duration of the mask can be redefined via UDP by using the external computer. 

The demodulator in the ADC module extracts the polarization signal 
from the masked stream $m(t_i)D(t_i)$ as follows: 
\begin{equation}\label{eq-demod}
 F_{\rm demod}(t_i) \equiv s^I_1(t_i)s^I_2(t_i)m(t_i)D(t_i),  
\end{equation}
where
\begin{numcases}{s^I_k(t_i)=} 
+1 & $0 \leq t_i < T_k/2$ \\ 
-1 & $T_k/2 \leq  t_i < T_k$ \\
s_k^I (t_i + nT_k), & 
\end{numcases}
$k=1,2$, $T_k=1/f_{{\rm c}k}$ and $n$ is an integer. 
Here, $s^I_k(t_i)$ is in-phase with the carrier $c_k(t_i)$. 
Summation is performed for all 6,400 points to form a 125~Hz ``Demod'' stream:
\begin{equation}
 {\rm Demod} \equiv \sum F_{\rm demod}(t_i). 
\end{equation} 
The extracted Demod stream is proportional to the Stokes $Q$ or $U$ parameters. 
Demodulation eliminates the $1/f$ noise below $f_{\rm c1}$. 

In case of quadratic phase demodulation (Quad), 
the Stokes $Q$ or $U$ parameters are also suppressed:
\begin{equation} 
   {\rm Quad} \equiv  \sum s^Q_1(t_i)s^I_2(t_i)m(t_i)D(t_i), 
\end{equation}
where $s^Q_1(t_i)$ is in the quadratic phase with $c_1(t_i)$, i.e.,
\begin{numcases}{s^Q_1(t_i)=} 
+1 & $0 \leq t_i < T_1/4$ \\ 
-1 & $T_1/4 \leq  t_i <3 T_1/4$ \\
+1 & $3T_1/4 \leq t_i < T_1$ \\ 
s^Q_1 (t_i + nT_1). & 
\end{numcases}
It is useful to monitor the Quad stream,  
because it only contains the noise whose level is the same as in the Demod stream. 

A simple summation without demodulation is used to extract the total power (TP) corresponding to the Stokes parameter $I
(\propto R + L)$:
\begin{equation} 
  {\rm TP} \equiv \sum m(t_i)D(t_i), 
\end{equation}
in which the intrinsic noise terms ($N_1$ and $N_2$) are neglected. 

\begin{figure}[!t]
\centering
 \includegraphics[width=9cm]{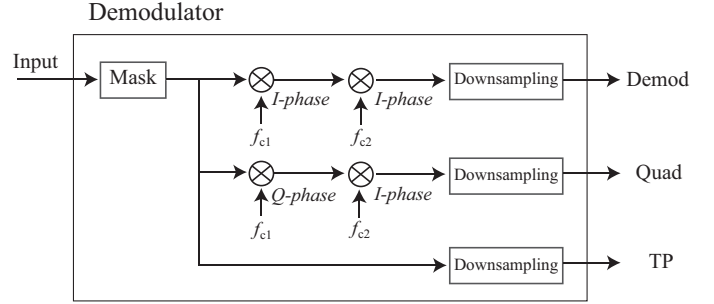}
 \caption{Demodulator logic is implemented for each input, i.e., 
 a single ADC module has 64 demodulators in total. 
 From a single input stream, the demodulator extracts the Demod, Quad, and TP streams simultaneously. }\label{fig-demod1}
\end{figure}

\begin{figure}
\centering
 \includegraphics[width=9cm]{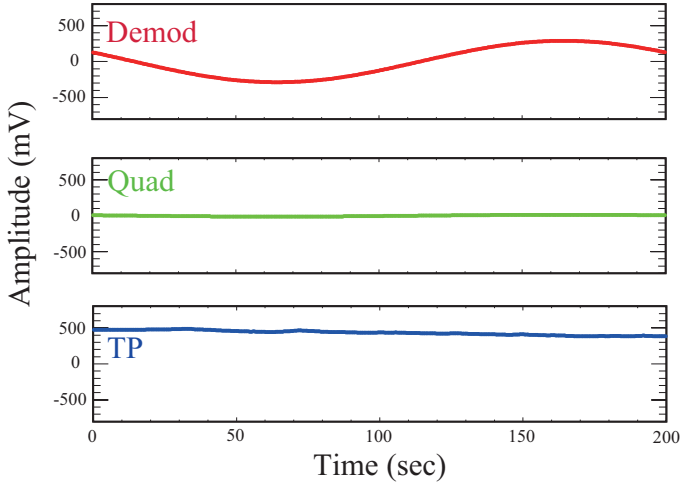}
 \caption{Demod, Quad and TP streams with a pseudo-polarization signal input. 
 The signal is appropriately extracted in each stream. 
 The input polarization is a sinusoidal wave. 
 The baseline drift as shown in the TP stream is completely suppressed in the Demod and Quad streams. }\label{fig-demod2}
\end{figure}

The demodulator is implemented in the FPGA for each input (Fig.~\ref{fig-demod1}). 
Using a pseudo-polarization signal input, we confirmed that the demodulator works appropriately, 
as shown in Fig.~\ref{fig-demod2}. 
The injected signal consists of a sinusoidal polarization signal (600~mV peak-to-peak amplitude at 5~mHz) 
and an offset that drifts at 400--500~mV.

\subsection{Intrinsic noise}
\begin{figure}
\centering
 \includegraphics[width=8cm]{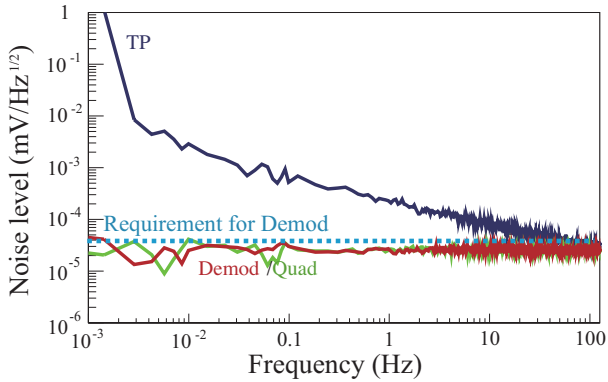}
 \caption{Noise spectra of TP, Demod and Quad with a $50~\Omega$ termination for the input. 
 The components of the $1/f$ noise are completely suppressed in the Demod and Quad spectra. 
 The white noise levels of the Demod and Quad spectra are $2 \times 10^{-5}$~mV/Hz$^{1/2}$, which fulfills the requirement.  
 }
\label{fig-noisespectra}
\end{figure}

\begin{figure}
 \includegraphics[width=8cm]{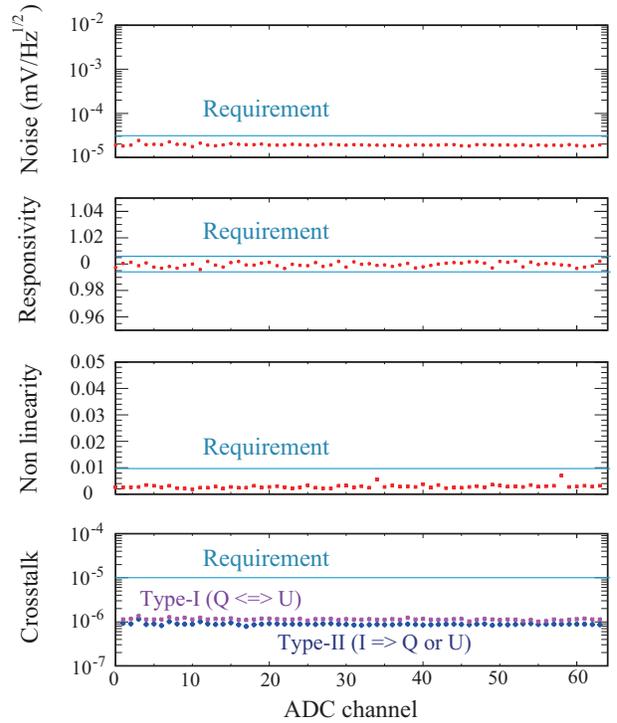}
 \caption{Performance of the ADC module as a function of the channel number. 
 From the top to the bottom panels; the white noise level, 
averaged responsivity, non linearity (maximum variation of responsivities with respect to the input signal level), 
 and two types of crosstalk among the channels. All requirements are satisfied.}
\label{fig-noiseForAllCh}
\end{figure}

We measure the intrinsic noise with a 50~$\Omega$ termination for each input.
Figure~\ref{fig-noisespectra} shows the noise spectra in one of the channels.
The $1/f$ components, which are obtained from the buffer amplifier, in the TP spectrum are completely suppressed in both Demod and Quad spectra.
The expected knee frequency of the polarimeter $1/f$ noise is approximately 1~kHz~\cite{QUIET-module}. 
Thus, $1/f$ noise suppression is well guaranteed. 
The measured noise floor in the Demod spectrum for each channel is shown in Fig.~\ref{fig-noiseForAllCh}.
We confirmed that the noise level is better than the requirements given in Sec.~\ref{sec-spec}.

\subsection{Linearity}
\begin{figure}
\centering
 \includegraphics[width=8cm]{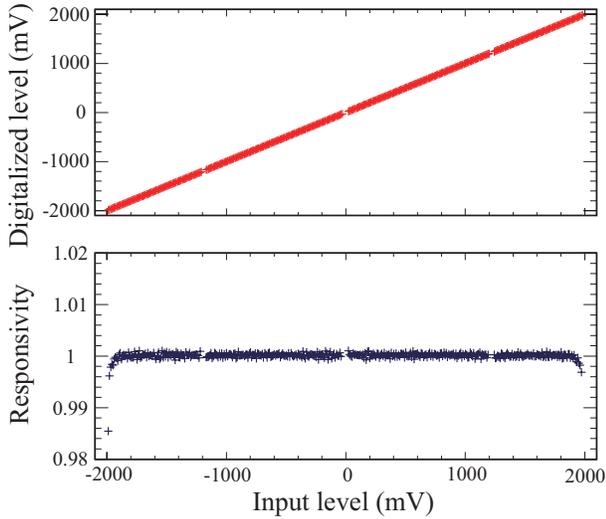}
 \caption{ADC response as a function of the input voltage (top panel). The responsivity for each input level is also shown (bottom panel). 
 We did not find any significant non linearity, i.e., variation in the responsivity, 
 for the range $-1,900$ to $1,900$~mV that we nominally use. 
 Because of the specification of the input signal injector, we could not cover the regions around 0~mV and $\pm 1,200$~mV for the input level. }
 \label{fig-adcline1}
\end{figure}

A linear response with respect to the input voltage level is required for the experiments. 
The top panel of Fig.~\ref{fig-adcline1} shows a typical response as a function of the input voltages. 
Responsivity as a function of the input voltages is also shown in the bottom panel of Fig.~\ref{fig-adcline1}. 
Within the $- 1,900$ to $1,900$~mV input range, we did not find any non linearity effects above $7 \times 10^{-3}$
(We nominally use the $- 1,900$ to $1,900$~mV range.). 
The averaged responsivity within the above range and non linearity as a function of channels are shown in the middle panels of Fig.~\ref{fig-noiseForAllCh}. 
Here, non linearity is defined as the difference between the maximum and minimum responsivities in the $- 1,900$ to $1,900$~mV range. 
We discovered slight non-uniformity among channels, i.e., $2 \times 10^{-3}$, in the standard deviations of averaged responsivity. 
Non linearity and non uniformity are lower than $10^{-2}$, which is the requirement to reduce a mimic $B$-mode intensity at $r=10^{-5}$. 

The requirement is determined when we do not have any calibration for the non linearity. 
We measure the Stokes parameters $Q$ and $U$ with different channels.
%
The polarization angle on the sky ($\phi$) is reconstructed with 
the measured responses for Stokes $Q$ and $U$,
\begin{equation}
 \phi = \frac{1}{2}\tan^{-1}\Biggl( \frac{a_U U}{a_Q Q}\Biggr),
\end{equation} 
where $a_Q$ and $a_U$ are the channel responsivities to measure $Q$ and $U$ at a given input power, respectively. 
The non linearities of $Q$ and $U$ and their non uniformity ($a_U \neq a_Q$) shifts the measured angle from the real value. 
Such an angle shift $\Delta \phi$ creates mimic $B$-modes \cite{komatsu}: 
\begin{equation}
 C_{\ell}^{BB,{\rm fake}}=C_{\ell}^{EE}\sin^2 (2\Delta \phi),
\end{equation}
where $C_{\ell}^{EE}$ is the $E$-mode power spectrum at the given angular wave-number $\ell$.  
We confirmed that the magnitude of non linearity and the sign and magnitude of the non uniformity of the responsivities are random, 
as shown in the middle panels of Fig.~\ref{fig-noiseForAllCh}.  
Therefore, the mimic $B$-modes are smeared with the square of the number of channels (2,000 channels). 
To determine the $B$-modes at $r=10^{-5}$ without any calibration for ADC responsivity, 
non linearity and non uniformity should be less than $10^{-2}$. 
 
\subsection{Crosstalk}

%
%
%
%
Two types of crosstalk among the channels must be considered: 
the crosstalk from one Demod stream to the other (type-I) and
the crosstalk from the TP stream to the Demod stream (type-II).
Type-I causes the Stokes $Q$ to $U$ crosstalk, which induces $E$-mode to $B$-mode mixing. 
In contrast, type-II causes Stokes $I$ to $Q$ (or $U$) leakage, i.e., 
spurious polarization from CMB temperature (unpolarized) anisotropy.
We require a crosstalk of less than $10^{-5}$ for either case, 
a requirement for suppressing the mimic $B$-modes to $r < 10^{-5}$.

Type-I crosstalk is measured by a pseudo-polarization signal with a 900~mV peak-to-peak amplitude at 0.2~Hz.
We inject the pseudo-polarization signal into one specific channel and measure the fluctuations in the other channels.
The bottom part of Fig.~\ref{fig-noiseForAllCh} shows the measurement results.
For type-II crosstalk, we inject a sinusoidal signal (800~mV peak-to-peak amplitude at 0.2~Hz) without any modulation into one specific channel. 
The measured fluctuations in other channels are also shown in the bottom part of Fig.~\ref{fig-noiseForAllCh}.
In both cases, we did not observe any significant crosstalk signal,
i.e., the measurements are saturated with the intrinsic noise levels.
We confirmed that the systematic bias possibly induced by the crosstalk is negligible.

\section{System test}\label{sec-systemtest}

\subsection{Timing synchronization}\label{sec-systest}
A single ADC module generates 768 byte (4 byte $\times$ 3 streams $\times$ 64 ADC chips) of demodulated data with the rate $f_{\rm record}=125$~Hz. 
Sixteen bytes of header information are added for each sample. 
The aggregate flow rate with the $32$ ADC modules is 25~Mbps. 

A 4 byte time counter is a component of the header. 
The counter is incremented by the edges of the 125~Hz carrier clock 
and it can be reset by a reset pulse. 
The pulse is controlled by the clock module. 
In the real observation, a pulse is sent at the beginning of continuous data taking. 
The typical duration of the data taking is approximately 1.5 hours.  
From the value of the time counter and the reset time calibrated with a GPS, 
we can find the time at which each data is demodulated.

\begin{figure}
\centering
 \includegraphics[width=7.5cm]{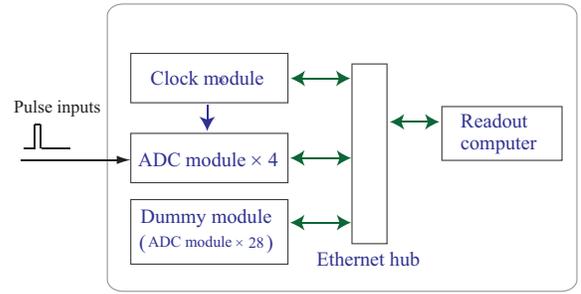}
 \caption{Configuration for the timing synchronization test. 
 The relationship between the data transfer and the system organization scheme is effectively reproduced by 
 the real situation in QUIET-II. 
 }\label{fig-adctest2}
\end{figure}

We tested timing synchronization among the components by applying a 25~Mbps load to a prototype system.
We constructed the system with one master clock module and four ADC modules
, one readout computer, and another computer as a ``dummy ADC module'' (Fig.~\ref{fig-adctest2}). 
They were physically connected via one Ethernet switch (D-Link, DES-1050G). 
The dummy ADC module sent 22~Mbps of dummy ADC data to the control computer using 28 different ports. 
Its data rate corresponds to that of the 28 ADC modules. 
In this setup, the situation in QUIET-II is effectively reproduced 
in terms of the data transfer and the system organization scheme relationship.  
Therefore, we virtually constructed the readout system for the QUIET-II experiment.

\begin{figure}
\centering
 \includegraphics[width=9cm]{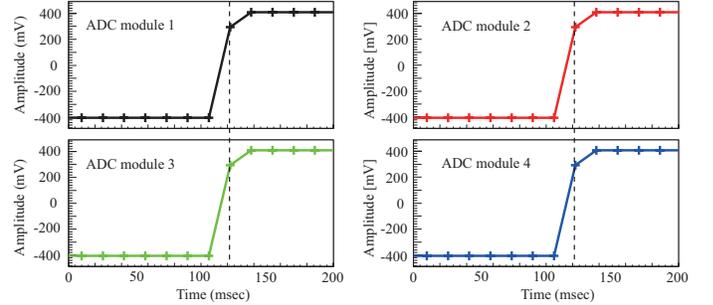}
 \caption{Total power response as a function of time for each ADC module.
Synchronization among the modules is confirmed with the edge
of the injected signals.}\label{fig-pulseinjection}
\end{figure}

We took data continuously for 16 hours with periodic (0.1 Hz) pulse signal injections into the ADC modules. 
We did not find any data loss or inconsistency among the time counters in the data. 
In addition, we confirmed timing synchronization among the ADC modules from the edge of the injected signals (Fig.~\ref{fig-pulseinjection}).


\subsection{Measurement of polarization with a polarimeter}\label{sec-read-polsignal}

\begin{figure}
\centering
 \includegraphics[width=9cm]{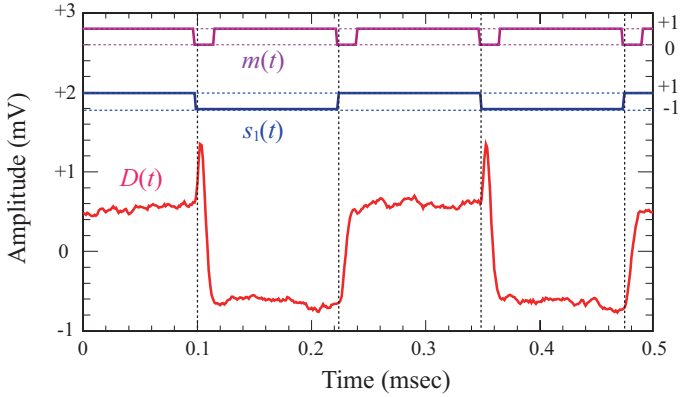}
 \caption{Digitized stream with $s_1(t)$ and $m(t)$. 
 The masking region is confirmed to be appropriate. }\label{fig-snapshot}
\end{figure}

By using a prototype polarimeter~\cite{QUIET-module2}, 
we examined the functions of the readout system, i.e., 
the mask for the ringing spikes and the extraction of the polarization signal. 
To study the mask, we analyzed the raw digitized data during $300~\mu$sec (see Sec.~\ref{subsec-demod}).  
For each sample, the data contain flags to indicate the 4~kHz carrier clock state $s_1(t_i)$ as well as the masking function $m(t_i)$ in the demodulator.
The measured response from the polarimeter as a function of time is shown in Fig.~\ref{fig-snapshot}.
The carrier clock state $s_1(t_i)$ and masking function $m(t_i)$ are also shown. 
We confirmed that phase mismatch occurs below the digitization interval, i.e., $< 1.25~\mu{\rm sec}$ (1/$f_{\rm A/D}$).
In addition, we validated the mask function for the ringing spikes. 

\begin{figure}
\centering
 \includegraphics[width=8cm]{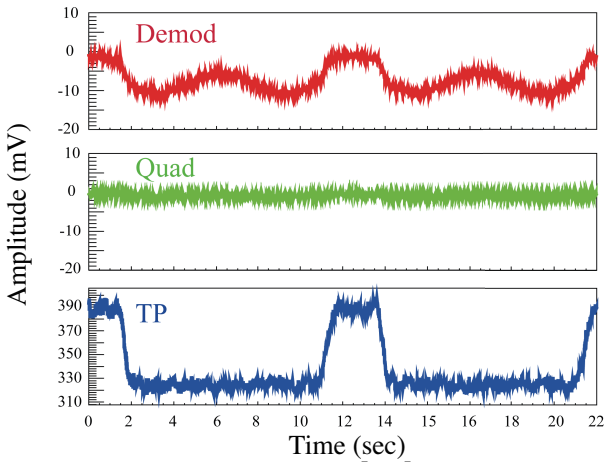}
 \caption{Extracted signals from the prototype polarimeter~\cite{QUIET-module2} by using the developed readout system. 
 The sinusoidal polarization signal and the periodic step-like variation for unpolarized radiation 
 (which is associate with the step-like offset variation for the polarization signal) are generated by the calibration system~\cite{hase}. 
 We observed the expected patterns in each stream. 
}\label{fig-hase}
\end{figure}

By using the calibration system~\cite{hase} with the prototype polarimeter and the developed system, 
the detection of the polarization signal was validated. 
When approximately 600~mK of polarization irradiated to the polarimeter, the
polarization angle was rotated by a frequency of about 0.1~Hz. 
The offset power (unpolarized radiation) also varied periodically
with the same rate.
The sinusoidal curve response with the baseline level shifts due to the
offset power variation were expected
(the cycle of the sinusoidal curve is twice the frequency of the rotation
because the polarization contains the tensor value).
We observed the expected response as shown in Fig.~\ref{fig-hase}.

\section{Conclusion}\label{sec-conclusion}
We developed a readout system for QUIET-II. 
The conceptual requirements (handling a large number of channels) are fulfilled by the master clock strategy, 
on-board demodulation functions, and end-to-end Ethernet connections. 
Physical compactness and logical simplicity enable us to further increase the polarimeters for future experiments.  
We demonstrated the extraction of the polarization signals from the polarimeter by using a prototype system with a configuration similar to that of the experiment. 
In long term operation, data loss and time counter mismatch were not observed. 
We validated that the system works according to the design. 
Therefore, the system is ready to be used in the QUIET-II experiment.

\appendix[Formulation for the polarimeter outputs]\label{app}
On the strip-line-coupled monolithic microwave integrated circuit devices, 
split right and left circular polarizations at the spectrum polarizer are amplified. 
Noises by HEMT amplifiers are added as follow: $g_1 E_R + n_1$ and $g_2 E_L + n_2$. 
Each phase switch modulate only the CMB signal components ($E_R$ and $E_L$). 
After the phase switches, two microwaves can be written as $c_1(t)g_1 E_R + n_1$ and $c_2(t)g_2 E_L + n_2$, respectively,  
where $c_1(t)$ and $c_2(t)$ are defined as the modulation terms in Sec.~\ref{sec-demod}. 
Consider only one diode that is sensitive to the Stokes $Q$ parameter; 
then the recombined microwave by the coupler is $1/\sqrt{2} (c_1(t)g_1 E_R + n_1 + c_2(t)g_2 E_L + n_2)$. 
The received power in the detector diode can be written as
\begin{equation}\label{eq-app}
 D(t) =  |\frac{1}{\sqrt{2}} (c_1(t)g_1 E_R + n_1 + c_2(t)g_2 E_L + n_2)|^2. 
\end{equation}
Assuming that polarimeter noises $n_1(t)$ and $n_2(t)$ are stationary Gaussian noises, i.e.,
$\langle n_1(t) \rangle$ = 0, $\langle n_2(t) \rangle$ = 0, and $\langle n_1(t)n_2(t) \rangle$ = 0, 
where $\langle X \rangle$ is the average of $X$ during each phase state of the primary modulation, 
Eq.~(\ref{eq-app}) is simplified as:
\setlength{\arraycolsep}{0.0em}
\begin{eqnarray}\label{eq-app1}
D(t_i)&{}={}& g_1 g_2 E_L E_R c_1(t_i)c_2(t_i) \IEEEnonumber \\
&&+ \frac{g_1^2E^2_L}{2} c_1(t_i)^2 + \frac{g_2^2E^2_R}{2} c_2(t_i)^2  \IEEEnonumber \\
&&+{} \frac{(n_1(t))^2}{2} |c_1(t_i)|^2 + \frac{(n_2(t))^2}{2} |c_2(t_i)|^2. 
\end{eqnarray}
\setlength{\arraycolsep}{5pt}
When $P = g_1 g_2Q/2$~~(here, $Q=2E_RE_L$), $L = g_1^2E^2_L/2$,  $R = g_2^2E^2_R/2$, $N_1(t_i)=(n_1(t_i))^2/2$, and $N_2(t_i) = (n_2(t_i))^2/2$, 
Eq.~(\ref{eq-app1}) will be transformed to Eq.~(\ref{eq-Dt}). 
By setting $c_2(t_i) = 1$, we can obtain Eq.~(\ref{eq-Dt0}).

\section*{Acknowledgment}
We are grateful to colleagues in University of Chicago for providing us 
with the ADC module design for the QUIET-I experiment. 
Special thanks to Professor Bruce Winstein who strongly encouraged our
development of the system.
We also acknowledge Jet Propulsion Laboratory and California Institute
of Technology for providing the QUIET-II polarimeter. 
We also thank Fermi National Accelerator Laboratory for providing several
electronics to control the polarimeter. 
We wish to thank Open Source Consortium of Instrumentation (OpenIt) for their cooperation and advice on the
electronics. 
We are grateful to Bee Beans Technologies Co., Ltd for its cooperation. 
This work was supported by MEXT and JSPS with a Grant-in-Aid for Scientific Research on Innovative Areas 21111002.

\ifCLASSOPTIONcaptionsoff
  \newpage
\fi



\bibliographystyle{IEEEtran.bst}

\bibliography{ADC_paper}

\begin{thebibliography}{10}
\providecommand{\url}[1]{#1}
\csname url@samestyle\endcsname
\providecommand{\newblock}{\relax}
\providecommand{\bibinfo}[2]{#2}
\providecommand{\BIBentrySTDinterwordspacing}{\spaceskip=0pt\relax}
\providecommand{\BIBentryALTinterwordstretchfactor}{4}
\providecommand{\BIBentryALTinterwordspacing}{\spaceskip=\fontdimen2\font plus
\BIBentryALTinterwordstretchfactor\fontdimen3\font minus
  \fontdimen4\font\relax}
\providecommand{\BIBforeignlanguage}[2]{{%
\expandafter\ifx\csname l@#1\endcsname\relax
\typeout{** WARNING: IEEEtran.bst: No hyphenation pattern has been}%
\typeout{** loaded for the language `#1'. Using the pattern for}%
\typeout{** the default language instead.}%
\else
\language=\csname l@#1\endcsname
\fi
#2}}
\providecommand{\BIBdecl}{\relax}
\BIBdecl

\bibitem{dodelson}
L.~Krauss, S.~Dodelson, and S.~Meyer, ``{Primordial Gravitational Waves and
  Cosmology},'' \emph{Science}, vol. 328, pp. 989--992, 2010.

\bibitem{ligo}
B.~Abbott \emph{et~al.}, ``{An Upper Limit on the Stochastic Gravitational-Wave
  Background of Cosmological Origin},'' \emph{Nature}, vol. 460, p. 990, 2009.

\bibitem{toba}
K.~Ishidoshiro \emph{et~al.}, ``Upper limit on gravitational wave backgrounds
  at 0.2 Hz with a torsion-bar antenna,'' \emph{Phys. Rev. Lett.}, vol. 106, p.
  161101, Apr 2011.

\bibitem{cassini}
J.~W. Armstrong, L.~Iess, P.~Tortora, and B.~Bertotti, ``{Stochastic
  gravitational wave background: Upper limits in the $10^{-6}$~Hz $10^{-3}$~Hz
  band},'' \emph{Astrophys. J.}, vol. 599, pp. 806--813, 2003.

\bibitem{QUIET}
QUIET collaboration ``{First Season QUIET Observations: Measurements of
  CMB Polarization Power Spectra at 43 GHz in the Multipole Range $25 \leq \ell
  \leq 475$},'' \emph{Astrophys. J.}, vol. 741, p. 111, 2011.

\bibitem{r}
S.~Weinberg, \emph{COSMOLOGY}.\hskip 1em plus 0.5em minus 0.4em\relax Great
  Clarendon Street, Oxford OX2 6DP: Oxford university press, 2008.

\bibitem{CAPMAP}
D.~Barkats \emph{et~al.}, ``First measurements of the polarization of the
  cosmic microwave background radiation at small angular scales from capmap,''
  \emph{Astrophys. J. Lett.}, vol. 619, no.~2, p. L127, 2005.

\bibitem{WMAP}
N.~Jarosik \emph{et~al.}, ``Design, implementation, and testing of the
  microwave anisotropy probe radiometers,'' \emph{Astrophys. J. Suppl.}, vol.
  145, no.~2, p. 413, 2003.

\bibitem{PLANK}
L.~Valenziano \emph{et~al.}, ``Planck-lfi: design and performance of the 4
  kelvin reference load unit,'' \emph{Journal of Instrumentation}, vol.~4,
  no.~12, p. T12006, 2009.

\bibitem{QUIET-module}
K.~A. Cleary, ``{Coherent polarimeter modules for the QUIET experimen},''
  \emph{Proc. SPIE}, vol. 7741, p. 77412H, 2011.

\bibitem{yuji}
Y.~Chinone, ``Measurement of cosmic microwave background polarization power
  spectra at 43 GHz with Q/U Imaging ExperimenT,'' Ph.D. thesis, Tohoku
  University, Sendai, JAPAN, February 2010.

\bibitem{nagai}
M.~Nagai \emph{et~al.}, ``Ethernet-based daq system for quiet-ii detectors,''
  \emph{Journal of Low Temperature Physics}, submitted for publication.

\bibitem{SiTCP}
T.~Uchida, ``Hardware-based tcp processor for gigabit ethernet,'' \emph{IEEE
  Transactions on Nuclear Science}, vol.~55, no.~3, pp. 1631--1637, 2008-06.

\bibitem{ADC}
``{ADC7674} data sheet,'' Analog devices, Norwood, MA 02062-9106, U.S.A.

\bibitem{QUIET-ADC1}
M.~Bogdan, ``Simultaneous sampling adc data acquisition system for the quiet
  experiment,'' in \emph{IEEE Nuclear Science Symposium Conference Record},
  2005, pp. 1077--1078.

\bibitem{QUIET-ADC2}
M.~Bogdan, D.~Kapner, D.~Samtleben, and K.~Vanderlinde, ``Digital control and
  data acquisition system for the quiet experiment,'' \emph{Nucl. Instr.
  Methods in Phys. Res. A}, vol. 572, no.~1, pp. 338--339, 2007.

\bibitem{komatsu}
E.~Komatsu \emph{et~al.}, ``Seven-year wilkinson microwave anisotropy probe
  (WMAP) observations: Cosmological interpretation,'' \emph{Astrophys. J.
  Suppl.}, vol. 192, no.~2, p.~18, 2011.

\bibitem{QUIET-module2}
R.~Reeves, ``Quiet coherent polarimeter modules,'' \emph{Journal of Low
  Temperature Physics}, submitted for publication.

\bibitem{hase}
M.~Hasegawa \emph{et~al.}, ``{Calibration System with Cryogenically-Cooled
  Loads for CMB Polarization Detectors},'' \emph{Rev. Sci. Instrum.}, vol.~81,
  p. 1054501, 2011.

\end{thebibliography}
\end{document}